**R3F: An R package for evolutionary dates, rates, and priors using the relative rate framework**

Running head: R3F: An R package for relative rate framework

Qiqing Tao[1,2], Sudip Sharma[1,2], Koichiro Tamura[3,4], and Sudhir Kumar[1,2*]

[1]Institute for Genomics and Evolutionary Medicine, Temple University
[2]Department of Biology, Temple University
[3]Department of Biological Sciences, Tokyo Metropolitan University, Hachioji, Tokyo, Japan
[4]Research Center for Genomics and Bioinformatics, Tokyo Metropolitan University, Hachioji, Tokyo, Japan

*Correspondence to:
    Sudhir Kumar
    Temple University
    Philadelphia, PA 19122, USA
    E-mail: s.kumar@temple.edu

**Abstract**

The relative rate framework (RRF) can estimate divergence times from branch lengths in a phylogeny, which is the theoretical basis of the RelTime method frequently-applied, relaxed clock approach for molecular dating that scales well for large phylogenies. The use of RRF has also enabled the development of computationally efficient and accurate methods for testing the autocorrelation of lineage rates in a phylogeny (CorrTest) and selecting data-driven parameters of the birth-death speciation model (ddBD), which can be used to specify priors in Bayesian molecular dating. We have developed *R3F*, an R package implementing RRF to estimate divergence times, infer lineage rates, conduct CorrTest, and build a ddBD tree prior for Bayesian dating in molecular phylogenies. Here, we describe *R3F* functionality and explain how to interpret and use its outputs in other visualization software and packages, such as MEGA, ggtree, and FigTree. Ultimately, *R3F* is intended to enable the dating of the Tree of Life with greater accuracy and precision, which would have important implications for studies of organism evolution, diversification dynamics, phylogeography, and biogeography.

**Availability and Implementation**: The source codes and related instructions for installing and implementing *R3F* are available from GitHub (https://github.com/cathyqqtao/R3F).

## 1. Introduction

The relative rate framework (RRF) is a system of equations that estimate relative rates among lineages exclusively from branch lengths in a phylogeny (Tamura et al. 2018). A lineage emanating from a node consists of the branch connecting to that node (stem) and all the branches in the descendant clade (**Fig. 1A**). In RRF, the principle of minimum rate change among lineages and their descendants is applied, enabling an algebraic solution for estimating relative times and lineage rates using branch lengths (Tamura et al. 2018). RRF is the theoretical foundation of the RelTime method, which is statistically accurate and computationally efficient in estimating divergence times in analyzing large computer-simulated and empirical datasets (review in Tao *et al.*, 2020). RelTime, as implemented in MEGA (Kumar et al. 2012; Tamura et al. 2021), has been cited in hundreds of research articles. However, all the functionalities of RRF are not available in a program that is easily conducive to scripting and programming environments. Therefore, we have developed the *R3F* package, in which RRF has been implemented in R for broader applications, including the use of use of RRF to generate a data-driven birth-and-death tree prior (ddBD) for Bayesian dating analyses (Tao et al. 2021). Using ddBD speciation prior in MCMCTree produces better time estimates than the default setting, especially when the number of calibrations is small (Tao et al. 2021). In addition, R3F offers functions to estimate dates and rates and a test for the presence of autocorrelation of branch rates in a phylogeny (CorrTest) (Tao et al. 2019). Knowledge of autocorrelation is interesting biologically and valuable when selecting a clock model in Bayesian dating analyses.

In the following, we describe our R implementations, input requirements, outputs produced, and interpreting results in the following. Because of their analytical nature, all the functions in *RRF* are ultra-fast, even for large phylogenies; they took less than one second for a phylogeny with 1000 tips (**Fig. 1B**).

## 2. Functions and implementations
### 2.1. Relative rate framework (RRF)

The *R3F* package requires a rooted phylogeny with branch lengths derived from molecular or non-molecular (e.g., morphological characters) sequences. The program accepts a text file containing a rooted tree in either Newick or NEXUS format. The users can also provide a file containing a list of tip names in the outgroup, which will be used internally to root the tree. Three primary RRF functions supplied in the *R3F* package are *rrf_rates, rrf_times*, and *rrf_rates_times*.

The *rrf_rates* function uses a phylogeny with branch lengths as input. It outputs a table of relative rates and a tree with rates in the Newick format. Note that RRF estimates lineage rates rather than branch rates. Lineage rates are assigned to the stem branches of the corresponding lineages (**Fig. 1A**) in the output Newick tree. Also, only the ingroup tips are included in the output because the equality of rates between ingroup and outgroup clades cannot be statistically tested (Kumar et al. 2016).

The *rrf_times* function has the same requirements and characteristics as *rrf_rates*, except it outputs a table of relative node times and a timetree in the Newick format. The third function, *rrf_rates_times*, computes relative rates and node times simultaneously. It outputs a table of rates and times and a timetree in the NEXUS format. The output tree can be plotted and branches colored based on the lineage rates with *plot==TRUE* in *rrf_rates_times* (**Fig. 1C**) or using other existing functions in other packages, such as *ggtree* (Yu et al. 2017). To examine the reliability of RRF functions, we analyzed 200 alignments that were simulated using an independent or autocorrelated rate model and a timetree of 446 species (Tamura et al. 2012). All three RRF functions produced relative rates and times very similar (slope > 0.979) to those produced by the

RelTime method, which uses RRF as the fundamental basis and is implemented in MEGA (Tamura et al. 2021), except for two alignments. For these alignments, trees contain too many zero-length branches (>30%). Because the treatments of zero-length branches in our implementation and RelTime are different, the differences between estimates are slightly larger (slope = 0.948 and 0.968 for two datasets). We also include a helpful tool, tree2table, for converting a tree in Newick or NEXUS of branch lengths, rates, or node times to a table containing node labels, branch labels, branch lengths, and node depths. This tool makes the direct comparison of multiple trees more convenient.

## 2.2. Functions based on the RRF framework

The *corrtest* function is included in the *R3F* package to classify the rate variation among branches and lineages in a phylogeny. It is based on a machine learning approach that uses the relative rates generated by RRF and produces a CorrScore that ranges from 0 to 1 (Tao et al. 2019). A high CorrScore is suggestive of autocorrelated rates. A *P*-value is also produced to assess the statistical significance of rejecting the independent rate model. The mean value and standard deviation of rates will be output if the user provides an anchor node time to convert relative node times to absolute node times. The other inputs required by the *corrtest* are a tree with branch lengths and the number of sister rate resampling replicates that are recommended for small phylogenies (<50 tips) (Tao et al. 2019).

The *ddbd* function estimates the Birth-Death (BD) speciation tree parameters before molecular dating in the MCMCTree (Yang, 2007). It uses relative times obtained using RRF (Tao et al. 2021). This function requires a tree with branch lengths along with one anchor time. An option is to estimate the proportion of species sampled (i.e., sampling fraction). If not user-provided, the *ddbd* function produces estimates of birth and death rates and sampling fractions. A figure showing the density distribution of node ages and the fitted BD curve will be plotted (**Fig. 1D**). Note that while the inferred BD parameters are helpful for molecular dating, these parameters are not direct estimates of birth and death rates because many combinations of parameter values can result in the same kernel density (Tao et al. 2021).

## 3. Conclusions

*R3F* will be a convenient tool for efficiently generating rates and times for small and large phylogenies. Its use will also facilitate a better selection of priors in Bayesian dating analysis. One can compare Bayesian estimates obtained using these priors with estimates obtained directly from RRF to evaluate the robustness of inferred times. Ultimately, the *R3F* package application can enable the Tree of Life's dating with greater accuracy and precision, which is important for inferring organism evolution, diversification dynamics, phylogeography, and biogeography studies.

## Acknowledgments

We thank Jose Barba-Montoya and Sara Vahdatshoar for testing the R package and providing many comments. This study was supported by a grant from the National Institutes of Health (R35GM139540-05) to S.K.

## Conflict of Interest

The authors declare no conflict of interest.

## Authors' contributions

S.K. and Q.T. conceived the idea; Q.T. and S.S. developed the package; Q.T., S.K., K.T., and S.S. wrote the manuscript.

## Data availability statement

The source code and example datasets are available from GitHub (https://github.com/cathyqqtao/R3F).

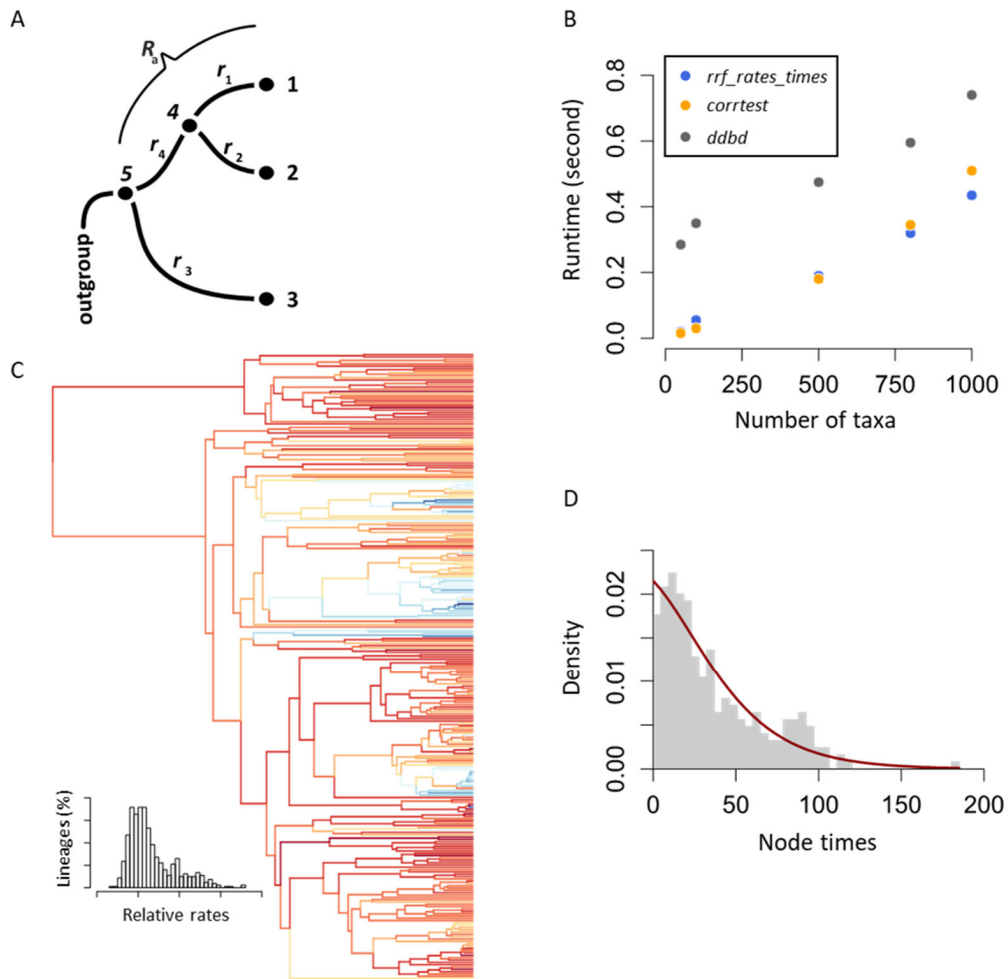

**Figure 1.**

(A) An example three-taxon phylogeny, rooted by an outgroup, with branch rates ($r$) and lineage rates ($R$) shown. The lineage rate $R_a$ includes the evolutionary rate of the stem branch ($r_4$) and the evolutionary rates of all the descendent branches ($r_1$ and $r_2$).

(B) Time taken by *rrf_rates_times*, *corrtest,* and *ddbd* functions for small and large phylogenies. Each point represents the average runtimes of two phylogenies: one simulated under the BD process with molecular rates drawn from an independent lognormal branch rate model, and the other simulated using the same BD process but with rates drawn from an autocorrelated branch rate model. All parameters used in the simulation were derived empirically.

(C) A timetree produced by *rrf_rates_times* in which branches are colored based on the value of relative rates; the inset shows the distribution of relative rates. For this analysis, branch lengths of the phylogeny shown were estimated using a sequence alignment from dos Reis *et al.* (2012). (D) The density distribution of node ages and the fitted curve used to generate BD parameters by *ddbd*.